# Influence of rotational motion of molecules on the thermal conductivity of solid $SF_6$, $CHCl_3$, $C_6H_6$, and $CCl_4$

## O.I. Purskii and N.N. Zholonko

*Technological State University of Cherkassy, 460 Shevchenko Blvd., Cherkassy 18006, Ukraine*
E-mail: pursky_O@ukr.net

## V.A. Konstantinov

*B. Verkin Institute for Low Temperature Physics and Engineering*
*of the National Academy of Sciences of Ukraine, 47 Lenin Ave., Kharkov 61103, Ukraine*
E-mail: konstantinov@ilt.kharkov.ua

The thermal conductivity of solid $SF_6$, $CHCl_3$, $C_6H_6$, and $CCl_4$ was investigated by the linear-flow method under saturated vapor pressures in the temperature range from 80 K to the corresponding melting temperatures and then recalculated for a constant density of the samples. The contributions of the phonon–phonon and phonon–rotation interactions to the total thermal resistance were separated using the modified method of reduced coordinates. It is shown that the phonon–rotation contribution to the thermal resistance of the crystals decreases as the rotational motion of the molecules attains more freedom.

## Introduction

Heat transfer in simple molecular crystals is determined by both translational and rotational motion of molecules at the lattice sites. As the temperature rises, the rotational motion of molecules in crystals can have basically the following stages: an increase in the libration amplitudes, jumplike reorientation of the molecules, increasing frequency of reorientations, hindered rotations of molecules and, finally, nearly free rotation of molecules. By choosing crystals with different parameters of the molecular interaction and varying the temperature it is possible to change the degree of orientational ordering and thus to investigate the influence of the rotational motion of molecules on the behavior of the thermal conductivity.

The objects of this study were the molecular crystals of $SF_6$, $CHCl_3$, $C_6H_6$, and $CCl_4$. The $SF_6$ molecule has octahedral symmetry. At 222.4 K sulphur hexafluoride crystallizes into a bcc lattice of $Im3m$ $(O_h^9)$ symmetry with two molecules per unit cell. As a result, the molecule and its surroundings have the same symmetry. On further cooling to 94.3 K a polymorphous transition occurs, which suppresses the symmetry of the translational and orientational subsystem to a monoclinic one, of space group $C2/m$

$(C_{2h}^3)$ with $Z = 6$, where $Z$ being coordination number. One-third of the $SF_6$ molecules take the high-symmetry $(2/m)$ positions and two-thirds of the molecules occupy the low-symmetry $(m)$ positions [1,2].

Sulphur hexafluoride is often classed with substances that have a plastic crystalline phase. Indeed, the relative molar entropy of melting $\Delta S_f / R$ of $SF_6$ is 2.61 [3], which is close to the Timmermans criterion. Here $R$ is universal gas constant. However, the nature of the orientational disorder in the high-temperature phase of $SF_6$ is somewhat different from that of plastic phases in other molecular crystals, where the symmetries of the molecule and its surroundings do not coincide. The interaction between the nearest neighbors in the bcc phase is favorable for molecular ordering caused by the S–F bonds along the {100} direction, and the interaction with the next nearest neighbors is dominated by repulsion between the F atoms. According to x-ray and neutron diffraction data [1,2,4] a strict order is observed in $SF_6$ (in phase I) just above the phase transition point. The structural dynamical factor $\Re$ characterizing the degree of the orientational order is close to unity in the interval 95–130 K. This feature sets off $SF_6$ from other plastic crystals, such as methane, carbon tetrachloride, adamantane and so on, where the long-range orientational order becomes disturbed immediately after the phase transition. Orientational disordering in $SF_6$ starts to intensify only above 140 K. As follows from the analysis of the terms of the Debye−Waller factor derived from neutron-diffractometric data for the high-temperature phase of $SF_6$, the F atoms have large effective libration amplitudes. As the temperature rises, the amplitudes increase to 20° and higher, but the F localization is still appreciable near the {100} direction. This implies that the orientational structure of $SF_6$ (I) does not become completely disordered even at rather high temperatures. The disordering itself is dynamic by nature. The increasing amplitudes of librations are not the only factor responsible for the increasing orientational disordering with rising temperature. It is, rather, connected with dynamic reorientations, which become more intensive due to frustrations of the molecular interactions.

Owing to these features, $SF_6$ offers a considerable possibility for investigating the influence of wide-range rotational states of the molecules on the thermal conductivity in a monophasal one-component system, where such states can vary from nearly complete orientational ordering to frozen rotation.

Chloroform ($CHCl_3$) has only one crystallographic modification in the whole interval of existence of the solid phase up to the melting temperature $T_m = 209.7$ K. It has the spatial symmetry $Pnma$ ($P_{2h}^{16}$) and four differently oriented molecules in the orthorhombic cell [5−7]. It is known from Raman and IR adsorption (20 K) data [7] that the translational modes take the frequency band up to 60 cm$^{-1}$ (86 K) and partially overlap the librational modes in the 60−100 cm$^{-1}$ band (86−144 K). The dipole moment of the $CHCl_3$ molecule is 1.01 D. Nuclear quadrupole resonance (NQR) on the $^{35}Cl$ nuclei has been observed in $CHCl_3$ up to the melting temperature [8]. These data indicate that there are no molecular reorientations at frequencies above $10^4$ s$^{-1}$. The high entropy of melting, $\Delta S_f / R = 5.4$, also attests to a high degree of ordering in $CHCl_3$ [3].

Solid benzene under the pressure of its own saturated vapor has only one crystallographic modification: it has the orthorhombic spatial symmetry $Pbca$ ($D_{2h}^{15}$) with four molecules per unit cell [9,10]. Benzene melts at 278.5 K and the melting-caused change in the entropy is $\Delta S_f / R = 4.22$ [3], which is much higher that the Timmermans criterion for *OD* phases. The high-temperature magnitude of the Debye temperature of $C_6H_6$ is 120 K [11].

In the interval 90−120 K the second NMR moment of $C_6H_6$ drops considerably as a result of the molecule reorientations in the plane of the ring around the six-fold axis [12]. The activation energy of the reorientational motion estimated from the spin−lattice relaxation time is 0.88 kJ/mole. The frequency of molecule reorientations at 85 K is $10^4$ s$^{-1}$. On a further rise of the temperature it increases considerably, reaching $10^{11}$ s$^{-1}$ near $T_m$. The basic frequency of the benzene molecule oscillations about the sixfold axis at 273 K is $1.05 \cdot 10^{12}$ s$^{-1}$ [13].

Carbon tetrachloride has an interesting feature: on cooling to 250.3 K liquid $CCl_4$ crystallizes into a face-centered cubic (fcc) form (Ia) with four molecules per unit cell; at several kelvins below 250.3 K it changes spontaneously to the rhombohedral phase (Ib), whose density is slightly higher, with 21 molecules per unit cell [14,15]. On a further cooling to 225.5 K, the rhombohedral phase transforms into a monoclinic one of space symmetry group $C2/c − (C_{2h}^6)$ with $Z = 32$. On heating, the low-temperature monoclinic phase (II) always changes to the rhombohedral form. Because of low entropy of melting, $\Delta S_f / R = 1.21$ [3], the phase (Ib) of $CCl_4$ may be classified as plastic.

The three forms of $CCl_4$ are quite closely related. The centers of mass of the molecules are only slightly shifted relative to their positions in the cubic and rhombohedral phases. Besides, the molecular orientations in the phase II correlate closely with the directions of the highest-density distribution function in the phase I.

According to experimental data, the character of the molecular motion in the plastic phase of $CCl_4$ is closely similar to that in the liquid state. For example, for $CCl_4$ no discontinuities are observed in the curve of spin−spin relaxation time $T_2$ of $^{35}Cl$ on plastic phase melting [16] and in the curve of reorientational correlation time obtained from Raman line broadening [17,18]. Zuk, Kiefte, and Clouter estimated the elastic constants of $CCl_4$ in the phase (I) by the Brillouin scattering method [19]. They discovered an anomalously high (as compared to solid inert gases) ratio of sound velocities in the <110> and <111> directions and interpreted this as an indication of a strong translation−orientation interaction.

## Experimental results and discussion

The thermal conductivity of solid $SF_6$, $CHCl_3$, $C_6H_6$, and $CCl_4$ was investigated by the linear-flow method under saturated vapor pressures in the temperature range from 80 K to the corresponding melting temperatures. A modified heat potentiometer was used [20], which permitted us to minimize the error in estimation of the thermal conductivity. The noncontrollable heat flows from thermal radiation were reduced considerably with a radiation shield on which the temperature field of the measuring cell was reproduced using a set of thermocouples and a precision heat controllers. The samples were grown from liquid and gaseous phases. At the bottom of the measuring ampoule the temperature was maintained close to that liquid $N_2$. The measurements were made on the two samples of each substance of 99.98% purity, the random error being within 5%. The results of measurement are shown in Figs. 1−4.

The isochoric thermal conductivity of the all four substances had been measured previously in narrow temperature intervals in the vicinity of the corresponding melting points [21−24]. The isobaric thermal conductivity of $C_6H_6$ and $CCl_4$ was also measured under pressure above 100 MPa in [25,26]. Our data are in good agreement with those results for the same conditions ($P,T$).

To find the correlation between experimental results and theory, it is reasonable to compare the data for constant volume and thus to exclude the influence of thermal expansion. The result obtained were recalculated for a constant density of the samples, whose molar volumes were: $V_m$, $cm^3/mole$: 58.25 ($SF_6$), 59.5 ($CHCl_3$), 70.5 ($C_6H_6$), 76.0 ($CCl_4$ (II)). This was done using the data for the volume dependence of the thermal conductivity [21−24] and thermal expansion [1,27]. The crystals had these volumes at the corresponding growth temperatures. The results were recalculated by the formula

$$\lambda_v = \lambda_p (V(T)/V_m)^g, \qquad (1)$$

where $\lambda_v$ and $\lambda_p$ are the isobaric and isochoric thermal conductivities, respectively; $V(T)$ is the current molar volume of the free sample; $V_m$ is the volume for which the results were recalculated; $g = -(\partial \ln \lambda / \partial \ln V)_T$ is the Bridgman coefficient.

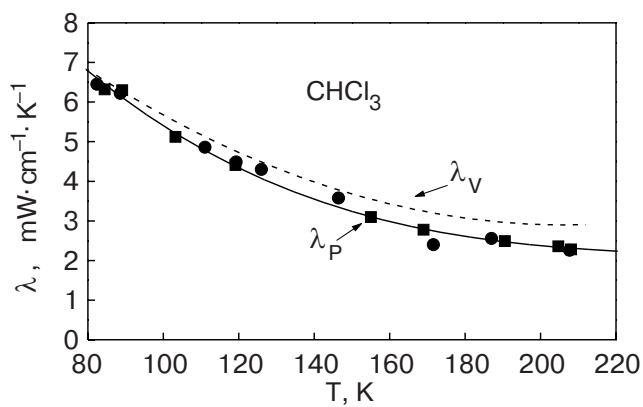

*Fig. 2.* The thermal conductivity (the solid line) of solid $CHCl_3$, measured under saturated vapor pressure. Rings and squares correspond to the two different samples. The dashed line is the thermal conductivity recalculated for the molar volume 59.5 $cm^3$/mole.

Figures 1−4 show the thermal conductivity measured under saturated vapor pressure (rings and solid lines) and the thermal conductivity recalculated for the corresponding molar volumes (broken lines). In chloroform, where the rotation of molecules is a pure librational motion and there are no reorientations, the isochoric thermal conductivity decreases was the temperature rises up to the melting point (Fig. 1). Similar behavior is observed in $SF_6$, $C_6H_6$, and $CCl_4$ (II) on the low-temperature side, with no intensive reorientations of the molecules. The isochoric thermal conductivity of $SF_6$, $C_6H_6$, and $CCl_4$ (II) passes through a minimum and then starts to grow. The minimum in the temperature dependence of the isochoric thermal conductivity occurs slightly above the temperature

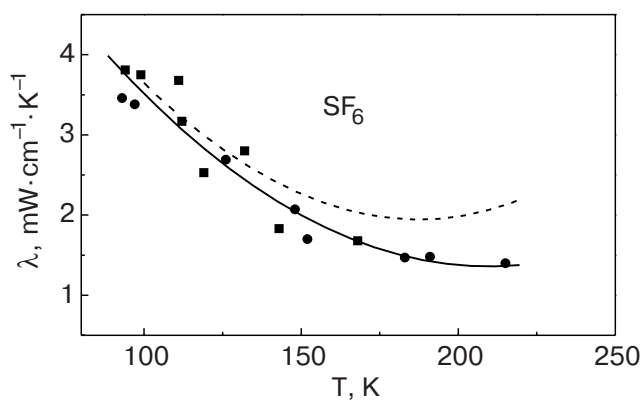

*Fig. 1.* The thermal conductivity (the solid line) of solid $SF_6$ in the high-temperature phase, measured under saturated vapor pressure. Rings and squares correspond to the two different samples. The dashed line is the thermal conductivity recalculated for the molar volume 58.25 $cm^3$/mole.

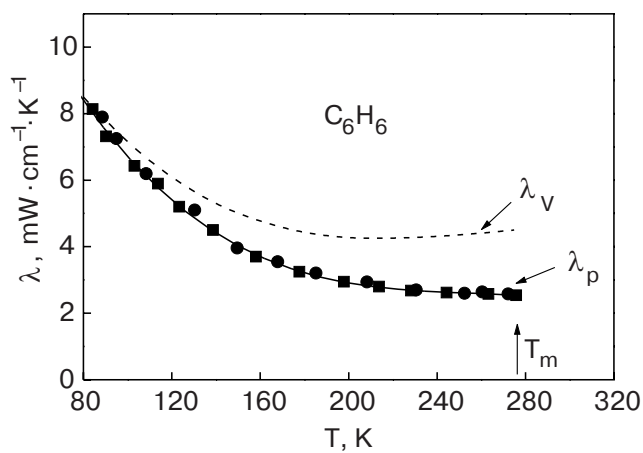

*Fig. 3.* The thermal conductivity (the solid line) of solid $C_6H_6$, measured under saturated vapor pressure. Rings and squares correspond to the two different samples. The dashed line is thermal conductivity recalculated for the molar volume 70.5 $cm^3$/mole.

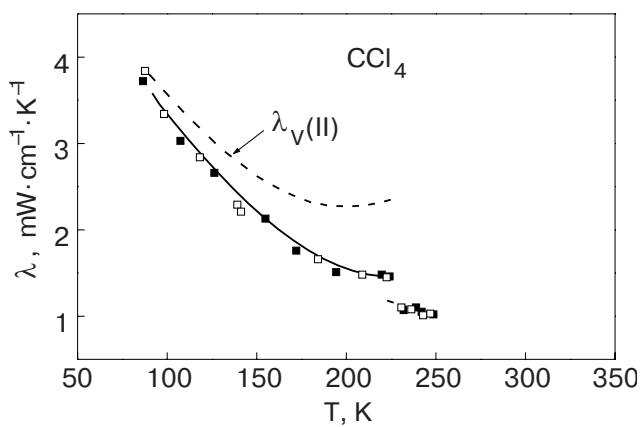

*Fig. 4.* The thermal conductivity (the solid line) of solid $CCl_4$ in the low-temperature phase, measured under saturated vapor pressure. Rings and squares correspond to the two different samples. The dashed line is the thermal conductivity recalculated for the molar volume 76.0 cm$^3$/mole.

at which intensive molecular reorientations begin in these crystals [1,12,16]. The minimum can therefore be attributed to the reorientational motion.

In this study the phonon–phonon and phonon–rotation contributions to the total thermal resistance were separated using a modified method of reduced coordinates [28]. It is important that with this method there is no need to involve any approximate model. As a rule, the reduction parameters are $T_{mol} = \varepsilon/k_B$, $\lambda_{mol} = k_B/\sigma^2\sqrt{\varepsilon/\mu}$, and $V_{mol} = N\sigma^3$, where $\sigma$ and $\varepsilon$ are the parameters of the Lennard–Jones potential, and $\mu$ is the molar weight, and $N$ is total number of particles. In this study the reductions parameters $T_{mol}$ and $V_{mol}$ are the temperatures and molar volumes of $SF_6$, $CHCl_3$, $C_6H_6$, and $CCl_4$ and solidified inert gases (krypton and xenon) at the critical points $T_{cr}$ и $V_{cr}$ [29–31] (see also Table).

Table

Reduced parameters, molar weight, and Bridgman coefficients for Kr, Xe, $SF_6$, $CHCl_3$, $C_6H_6$, and $CCl_4$

| Sub-stance | $T_{mol}$, K | $V_{mol}$, cm$^3$/mole | $\mu$ | $W_{mol}(1/\lambda_{mol})$, m·K/W | $g$ | Refe-rences |
|---|---|---|---|---|---|---|
| Kr | 209.4 | 92.01 | 83.8 | 8.06 | 10.2 | [29] |
| Xe | 289.7 | 119.4 | 131.3 | 10.0 | 9.6 | [29] |
| $SF_6$ | 318.7 | 201.45 | 146.05 | 13.51 | 5.2 | [22,30] |
| $CHCl_3$ | 536.6 | 238.8 | 119.4 | 10.9 | 4.0 | [21,31] |
| $C_6H_6$ | 562.0 | 260.0 | 78.1 | 9.43 | 7.5 | [23,30] |
| $CCl_4$ | 556.4 | 257.0 | 153.8 | 13.2 | 6.0 | [24,31] |

The reason for this choice of parameters is as follows. For simple molecular substances $T_{cr}$ and $V_{cr}$ are proportional to $\varepsilon$ and $\sigma^3$, respectively. However, the accuracy of the critical parameters is much higher than that of the binomial potential parameters. Note that $\sigma$ and $\varepsilon$ are essentially dependent on the choice of binomial parameter and the method of its determination. The phonon–phonon and phonon–rotation components of the thermal resistance can be separated assuming that (i) the total thermal resistance $W = 1/\lambda$ of simple molecular crystals is a sum of phonon–phonon $W_{pp}$ and phonon–rotation $W_{pr}$ contributions: $W = W_{pp} + W_{pr}$, and (ii) in reduced coordinates ($W^* = W/W_{mol}$, $T^* = T/T_{mol}$) the component resulting from phonon–phonon scattering $W_{pp}$ is identical to that in solid inert gases at equal reduced molar volumes $V^* = V/V_{mol}$.

The calculation results are shown in Figs. 5–8. The phonon–phonon component of the thermal resistance $W_{pp}$ is practically (to within 2–3%) independent of the inert gas chosen for comparison. In solid $CHCl_3$ (Fig. 6) the thermal resistance $W_{pr}$ can be attributed to extra phonon scattering by collective rotational excitations whose density increases as the temperature rises. This is in good agreement with the data in [8], that suggests complete orientational ordering in solid $CHCl_3$ persisting up to the melting temperature. The extra contribution to the thermal resistance from the rotational degrees of freedom of the molecules is 80% of the phonon–phonon component. Unlike solid $CHCl_3$, in which the phonon–rotation component of the thermal resistance increases with growing temperature, in solid $SF_6$, $C_6H_6$, and $CCl_4$ the translation–rotational thermal resistance increases with temperature, passes through a maximum and then starts to decrease. The effect may be due to attenuation of phonon scattering by collective rotational excitations

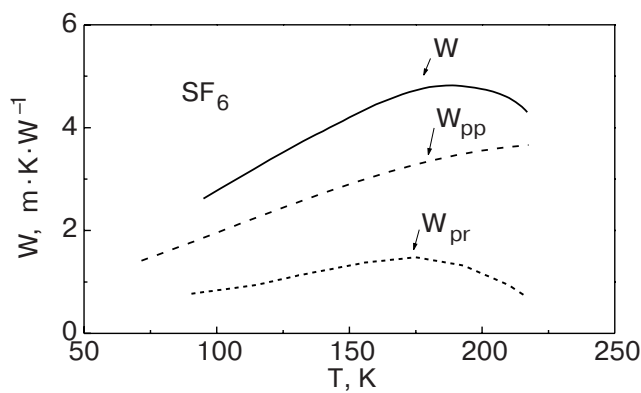

*Fig. 5.* Contributions of phonon–phonon scattering $W_{pp}$ and phonon scattering caused by rotational molecular excitations $W_{pr}$ to the total thermal resistance $W$ of solid $SF_6$ with the molar volume 58.25 cm$^3$/mole.

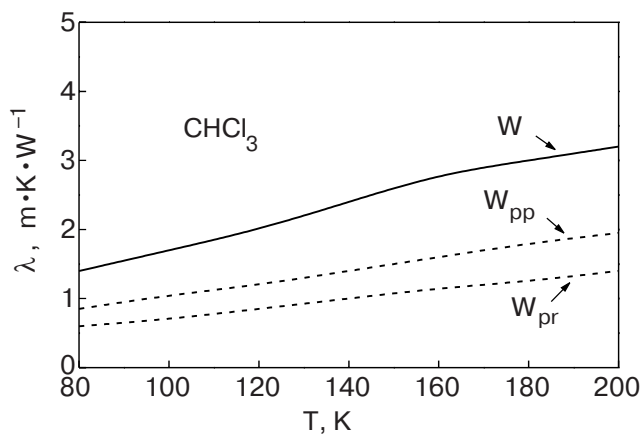

*Fig. 6.* Contributions of phonon−phonon scattering $W_{pp}$ and phonon scattering caused by rotational molecular excitations $W_{pr}$ to the total thermal resistance $W$ of solid $CHCl_3$ with the molar volume 59.5 cm³/mole.

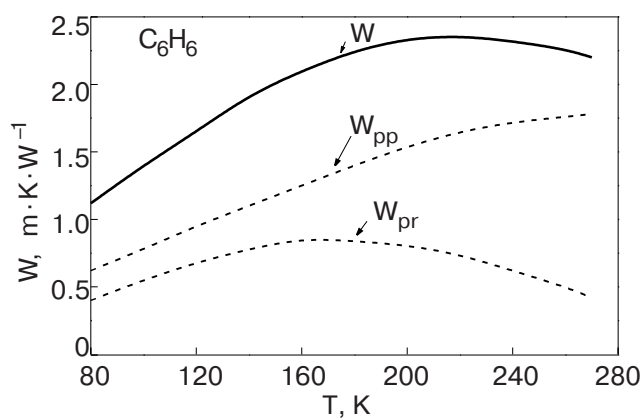

*Fig. 7.* Contributions of phonon−phonon scattering $W_{pp}$ and phonon scattering caused by rotational molecular excitations $W_{pr}$ to the total thermal resistance $W$ of solid $C_6H_6$ with the molar volume 70.5 cm³/mole.

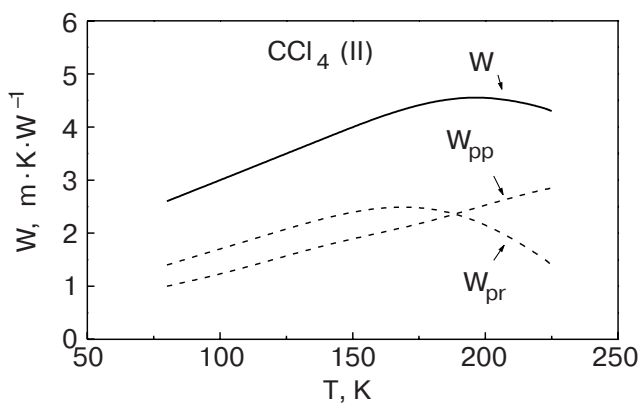

*Fig. 8.* Contributions of phonon−phonon scattering $W_{pp}$ and phonon scattering caused by rotational molecular excitations $W_{pr}$ to the total thermal resistance $W$ of solid $CCl_4$ with the molar volume 76.0 cm³/mole.

as the correlation of the neighboring molecular rotations decreases.

## Conclusions

For the example of the simple molecular crystals $SF_6$, $CHCl_3$, $C_6H_6$, and $CCl_4$ the correlation between the rotational motion of molecules at the lattice sites and the behavior of thermal conductivity has been investigated. It is shown that the isochoric thermal conductivity increases as the frequency of reorientations grows higher with rising temperature. It is found that this effect is connected with phonon scattering by collective rotational excitations, which attenuates was the rotational correlations of the neighboring molecules become weaker.